\documentclass[print]{raa}

\usepackage{graphicx,times}
\usepackage{natbib}
\usepackage{floatrow}
\usepackage{amssymb,amsmath}
\bibpunct{(}{)}{;}{a}{}{,}

\usepackage[a4paper=true,dvipdfm=true,pagebackref=true]{hyperref}
\hypersetup{colorlinks = true, linkcolor = green, anchorcolor = red, citecolor = blue, filecolor = red, pagecolor = red, urlcolor = red}

\begin{document}

   \title{Modeling the fast optical transient SN 2019bkc/ATLAS19dqr with a central engine and implication for its origin
%\footnotetext{$*$ Supported by the National Natural Science Foundation of China.}
}
%   \subtitle{I. Place Your Subtitle Here}

   \volnopage{Vol.0 (20xx) No.0, 000--000}      %%preserved for Editor. DOn't remove!
   \setcounter{page}{1}          %%starting page, preserved for Editor. DOn't remove!

   \author{Jian-He Zheng and Yun-Wei Yu\mailto{yuyw@mail.ccnu.edu.cn}}
   \institute{$^1$Institute of Astrophysics, Central China Normal University, Wuhan 430079, China\\
   $^2$Key Laboratory of Quark and Lepton Physics (Central
China Normal University), Ministry of Education, Wuhan 430079,
China}
   \email{yuyw@mail.ccnu.edu.cn}

\date
{\small Received~~20xx month day; accepted~~20xx~~month day}

\abstract{
Modern wide-field high-cadence surveys have revealed the significant diversity of optical transient phenomena in their luminosity and timescale distributions, which led to the discovery of some mysterious fast optical transients (FOTs). These FOTs can usually rise and decline remarkably in a timescale of a few days to weeks, which are obviously much rapider than ordinary supernovae. SN 2019bkc/ATLAS19dqr is one of the fastest detected FOTs so far and, meanwhile, it was found to be un-associated with a host galaxy. These discoveries provide a good chance to explore the possible origins of FOTs. So, we model the light curves of SN 2019bkc in details. It is found that SN 2019bkc can be well explained by the thermal emission of an explosion ejecta that is powered by a long-lasting central engine. The engine could be a spinning-down millisecond magnetar or a fallback accretion onto a compact object. Combining the engine property, the mass of the ejecta, and the hostlessness of SN 2019bkc, we suggest that this FOT is likely to originate from a merger of a white dwarf and a neutron star.
}

   \authorrunning{J.-H. Zheng $\&$ Y.-W. Yu }            %author_head in even pages
   \titlerunning{Modeling SN 2019bkc/ATLAS19dqr and implication for its origin}  % title_head in odd pages

   \maketitle

\section{Introduction}           %% first-level sections will be auto-capitalized
\label{sect:intro}
Astronomical transient phenomena, particularly, the ones occurring in extragalactic galaxies usually indicate an extremely huge energy release as a result of catastrophic collapses of massive stars or binaries. Specifically, supernovae (SNe) are undoubtedly the most representative of such transients, which are the primary targets of many current wide-field surveys. Thanks to these modern surveys with their tight cadences, a huge diversity has appeared in the distribution of all SN-like transients in the space of peak luminosity and emission timescale. It is then discovered that
some unusual fast optical transients (FOTs) can rise and
decline significantly from view in a few days or weeks, which is much more rapider than the typical SNe \citep{Drout2014ApJ...794...23D,Prentice2018ApJ...865L...3P,Pursiainen2018MNRAS.481..894P,Perley2019MNRAS.484.1031P,McBrien2019ApJ...885L..23M,Chen_2020,Gillanders2020MNRAS.497..246G,Prentice2020A&A...635A.186P}. Such fast evolving behaviors indicate that these FOTs
probably have origins very and even intrinsically different from normal SN explosions. In any case, within the basic framework of the stellar explosion, the short duration of FOTs could indicate that their progenitors are compact objects or ultra-stripped stars, because the explosion of these stars can naturally produce a low-mass ejecta of a short photon diffusion timescale \citep{Yu_2015}.

The most famous FOT is undoubtedly the kilonova AT2017gfo, which was discovered in the follow-up observations of the gravitational wave (GW) event on 17 August 2017 and had played a crucial role in localizing and identifying the origin of the GW signal \cite[NSs; ][]{2017PhRvL.119p1101A,Andreoni2017PASA...34...69A,Arcavi2017Natur.551...64A,Chornock2017ApJ...848L..19C,Coulter2017Sci...358.1556C,Cowperthwaite2017ApJ...848L..17C,Drout2017Sci...358.1570D,Evans2017Sci...358.1565E,
Kasliwal2017Sci...358.1559K,Lipunov2017ApJ...850L...1L,Nicholl2017ApJ...848L..18N,Pian2017Natur.551...67P,Smartt2017Natur.551...75S,Tanvir2017ApJ...848L..27T,Troja2017Natur.551...71T,Utsumi2017PASJ...69..101U,
Valenti2017ApJ...848L..24V}.
Such kilonova emission was first predicted by \cite{Li1998ApJ...507L..59L} and elaborately described by \cite{Metzger2010MNRAS.406.2650M}, as a promising electromagnetic counterpart of a GW event. Following these pioneering works, it is widely suggested that the AT2017gfo emission can be powered by the radioactive decay of heavy r-process elements synthesized during the merger of compact objects \citep{Kasen2017Natur.551...80K,Metzger2017LRR....20....3M}. However, from the detailed modeling of the AT2017gfo data, it can be found that the preconceived radioactive explanation is actually very questionable \citep{LiSZ2018ApJ...861L..12L}. Instead, an extra energy source is necessarily required, which can even be dominant over the radioactive power \citep{Yu2018ApJ...861..114Y,LiSZ2018ApJ...861L..12L}. Specifically, such an extra energy source can be provided by a remnant massive NS formed from the merger, as previously suggested by \cite{Yu2013ApJ...776L..40Y} and \cite{Metzger2014MNRAS.439.3916M}.
The existence of such a post-merger massive NS in the GW170817 event has further been supported by the works of \cite{Piro2019MNRAS.483.1912P} and \cite{Ren2019ApJ...885...60R}.

Fairly speaking, it is not very surprising to find a long-lasting energy engine from an FOT, since such an engine has been widely used to explain some extreme transient phenomena (see \cite{2019AIPC.2127b0024Y} for a brief review), such as superluminous SNe \citep{Woosley2010ApJ...719L.204W,Kasen_2010,Dexter_2013} and gamma-ray bursts \citep{Dai1998a,Dai1998b,Dai2004ApJ...606.1000D,Dai20061127,Yu2010ApJ...715..477Y}. Just following this knowledge, it has been previously suggested by \cite{Yu_2015} that, at least, the FOTs of an ultrahigh luminosity are very likely to be powered by a central engine, where the luminosity is too high to be explained by the radioactive scenario even though all of the explosively-ejected material is supposed to be radioactive. Generally, the nature of the central engine of an FOT could be a spinning-down neutron star (NS) or a fallback accretion onto a compact object, which can evolve from different progenitors.

Recent years, the implementation of modern surveys and the discovery of AT 2017gfo has effectively promoted the discovery of a lot of luminous FOTs \citep{McBrien2019ApJ...885L..23M}, among which SN 2019bkc/ATLAS19dqr is the most rapidly declining one that was discovered by Asteroid Terrestrial-impact Last Alert System \cite[ATLAS; ][]{Tonry_2018,Chen_2020}. This source is the focus of this paper, because of
its unprecedented magnitude decline after the peak. In about four days, the luminosity of SN2019bkc decayed by a half. This decline rate is very close to the situation of AT2017gfo \citep{Chen_2020,Prentice2020A&A...635A.186P}. In principle, such a rapid evolution can appear in some ultra-stripped SNe \citep{Tauris2013ApJ...778L..23T} or in the shock breakout emission of some normal or failed SNe, which are however disfavored by the un-association of SN 2019bkc with a host galaxy. Therefore, following the considerations in \cite{Yu_2015} and \cite{Yu2018ApJ...861..114Y}, here we would like to connect SN 2019bkc with a compact object progenitor and model its temporal evolution with a central engine. It is expected that an implication for the origin of SN 2019bkc could be found from the properties of the engine and the explosion ejecta.

\section{Modeling the temporal evolution of SN 2019bkc}
\label{sect:Mod}
\subsection{Model description}
For a hot explosion ejecta of a mass $M_{\rm ej}$ and a radius $R$, the bolometric
luminosity of its thermal emission, which is determined by the heat diffusion in it, can be roughly estimated by \citep{Kasen_2010,Kotera2013MNRAS.432.3228K,Yu_2015}
\begin{eqnarray}
L_{\rm bol}\sim\frac{cE_{\rm int}}{{R_{\rm }\tau}}(1-e^{-\tau}),
\end{eqnarray}
where $c$ is the speed of light, $E_{\rm int}$ is the internal energy of the ejecta, and $\tau=3\kappa M_{\rm ej}/4\pi R^2$ is the optical depth with $\kappa$ being the opacity. This expression combines two asymptotic properties of the emission as $L_{\rm bol}= cE_{\rm int}/( R\tau)$ for $\tau\gg 1$ and
$L_{\rm bol}= cE_{\rm int}/R$ for $\tau \ll1$.
The evolution of the internal
energy is simultaneously determined by the energy conversation as
\begin{eqnarray}
\frac{ d E_{\rm int}}{d t} =  L_{\rm ce} +L_{\rm rad}- L_{\rm bol}- 4\pi R_{\rm
}^2 pv_{\rm }  ,\label{Eint}
\end{eqnarray}
where $L_{\rm ce}$ and $L_{\rm rad}$ are the heating rate due to the central engine and the radioactivity, respectively,
$p=\frac{1}{3}(E_{\rm int}/{\frac{4}{3}}\pi R_{\rm }^3)$ is the radiation-dominated pressure, and $v$ is the expansion velocity of the ejecta which determines the ejecta radius by $dR=vdt$. The expansion velocity can in principle be increased, because of the increase of the kinetic energy $E_{\rm k}$ of the ejecta through the work by the pressure.

For the radioactive power, if it is dominated by the decay of nickels as usual as for typical SNe, then the corresponding heating rate can be given by \citep{Colgate1980ApJ...237L..81C,Arnett1980}
\begin{equation}
    L_{\rm rad,Ni}=M_{\rm Ni}[(\epsilon_{\rm Ni}-\epsilon_{\rm Co})e^{-t/\tau_{\rm Ni}} + \epsilon_{\rm Co}e^{-t/\tau_{\rm Co}} ],
\end{equation}
where $M_{\rm Ni}$ is the total mass of $^{56}$Ni, $\epsilon_{\rm Ni}=3.90\times10^{10}{\rm erg~s^{-1}g^{-1}}$ and $\epsilon_{\rm Co}=6.78\times10^{9}{\rm erg~s^{-1}g^{-1}}$, and $\tau_{\rm Ni}=8.76$ days and $\tau_{\rm Co}=111.42$ days are the lifetimes of the radioactive elements. Alternatively, if the dominant radioactive elements are r-process elements as for typical kilonovae, then we should employ \citep{Korobkin2012MNRAS.426.1940K}
\begin{equation}
L_{\rm rad,R}=4\times10^{18}M_{\rm ej}\left[{\frac{1}{2}}-{\frac{1}{\pi}}{\rm
arctan}\left({\frac{{t-t_0}}{\sigma}}\right)\right]^{1.3}\rm
erg~s^{-1},\label{Lrpro}
\end{equation}
with $t_0=1.3$ s and $\sigma=0.11$ s. Meanwhile, the central engine of the FOT can also have two representative natures including (i) a spinning-down NS, which is usually considered to spin at a near-Keplerian frequency and be highly magnetized, and (ii) an accretion of fallback material onto the central compact object. In the former case, the spin-down luminosity of the NS can be estimated by its magnetic dipole radiation power as usual as \citep{1983bhwd.book.....S}
\begin{eqnarray}
L_{\rm sd}=L_{\rm sd,i}\left(1+{\frac{t}{t_{\rm sd}}}\right)^{-2}.
\label{spindown}
\end{eqnarray}
In the later case, as usual, we assume simply the accretion luminosity to be proportional to the accretion
rate and then have \citep{Piro2011ApJ...736..108P}
\begin{eqnarray}
L_{\rm ac}=L_{\rm ac,\max}\left[\left({\frac{t}{t_{\rm
ac}}}\right)^{-1/2}+\left({\frac{t}{t_{\rm
ac}}}\right)^{5/3}\right]^{-1}
\label{fallback}.
\end{eqnarray}
Here, the characteristic luminosity $L_{\rm sd,i}$ or $L_{\rm ac,\max}$ and timescale $t_{\rm sd}$ or  $t_{\rm ac}$ are taken as free parameters in our calculations. It should be pointed out that, when we use $L_{\rm sd}$ or $L_{\rm ac}$ to determine the energy injection rate $L_{\rm ce}$ in Equation (\ref{Eint}),  an extra factor of $\xi$ should be multiplied, the value of which is in principle determined by the thermalization efficiency of the energy and also influenced by the possible anisotropic distribution of the energy outflow.

After the bolometric luminosity of the FOT is given, a black-body effective temperature can be given by
\begin{eqnarray}
T_{\rm BB}=\left(\frac{L_{\rm bol}}{4\pi R_{\rm ph}^2\sigma}\right)^{1/4},
\end{eqnarray}
which is crucial for determining the radiation spectrum and then the chromatic luminosities for different filters, where $\sigma$ is the Stefan-Boltzmann constant and $R_{\rm ph}$ is the photospheric radius.
Following \citet{Arnett1982ApJ...253..785A}, we define the photospheric radius by
\begin{equation}
R_{\rm ph}=R_{\rm }-\frac{2}{3}\lambda,\label{rph}
\end{equation}
where $\lambda(t)=1/\kappa\rho(t)$ is the mean free path of photons.
Obviously, the motion of the photosphere due to the expansion of the ejecta is dependent on the material distribution in the ejecta (see \citealt{Liu_2018} for a general investigation). For simplicity, we adopt the single power-law density profile in our calculations.
\begin{equation}
    \rho(r,t)=\rho(R_{\rm i},0)\left[\frac{R_{\rm i}-R_{\rm min,i}}{R_{\rm}(t)-R_{\rm min}(t)}\right]^3 x^{\rm -\delta},
\end{equation}
where $R=R_{\rm i}+v_{\rm}t$ and $R_{\rm min}=R_{\rm min,i}+v_{\rm min}t$ are the outmost and inmost radius of the ejecta with $R_{\rm i}$ and $R_{\rm min,i}$ being their initial values, the dimensionless radius is defined as $x\equiv(r-R_{\min})/(R-R_{\min})$.
For such a velocity distribution, the relationship between the leading velocity and the kinetic energy can be written as $v_{\rm
}=(2I_{\rm M}E_{\rm k}/I_{\rm K}M_{\rm ej})^{1/2}$ with $I_{\rm M} =\frac{1}{3-\delta}$ and $I_{\rm K} =\frac{1}{5-\delta}$.
Finally, when the value given by Eq. (\ref{rph}) is smaller than $R_{\min}$, we artificially take $R_{\rm ph}=R_{\min}$.

\subsection{Fitting results}

\begin{figure}
    \centering
    \includegraphics[scale=0.6]{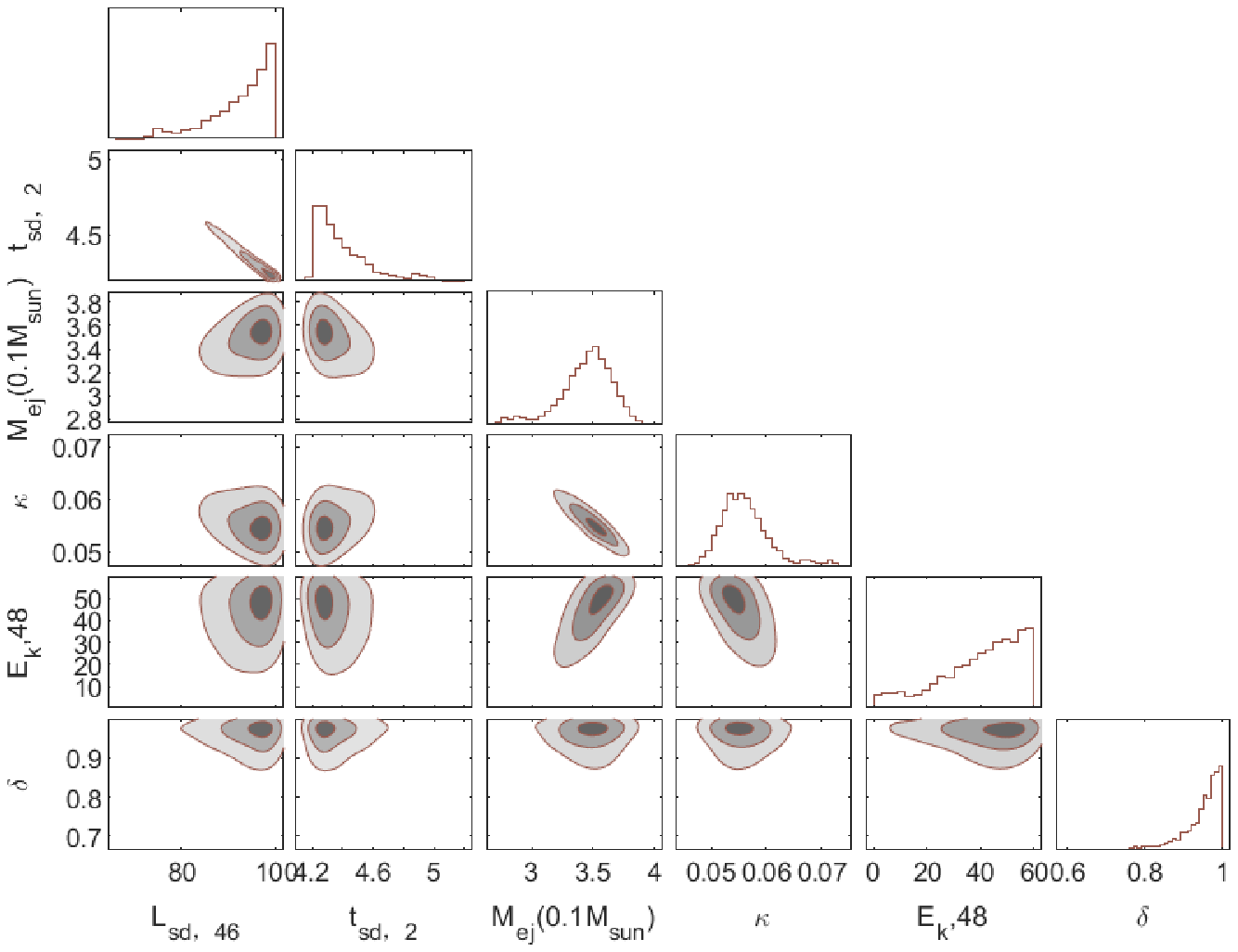}
    \includegraphics[scale=0.6]{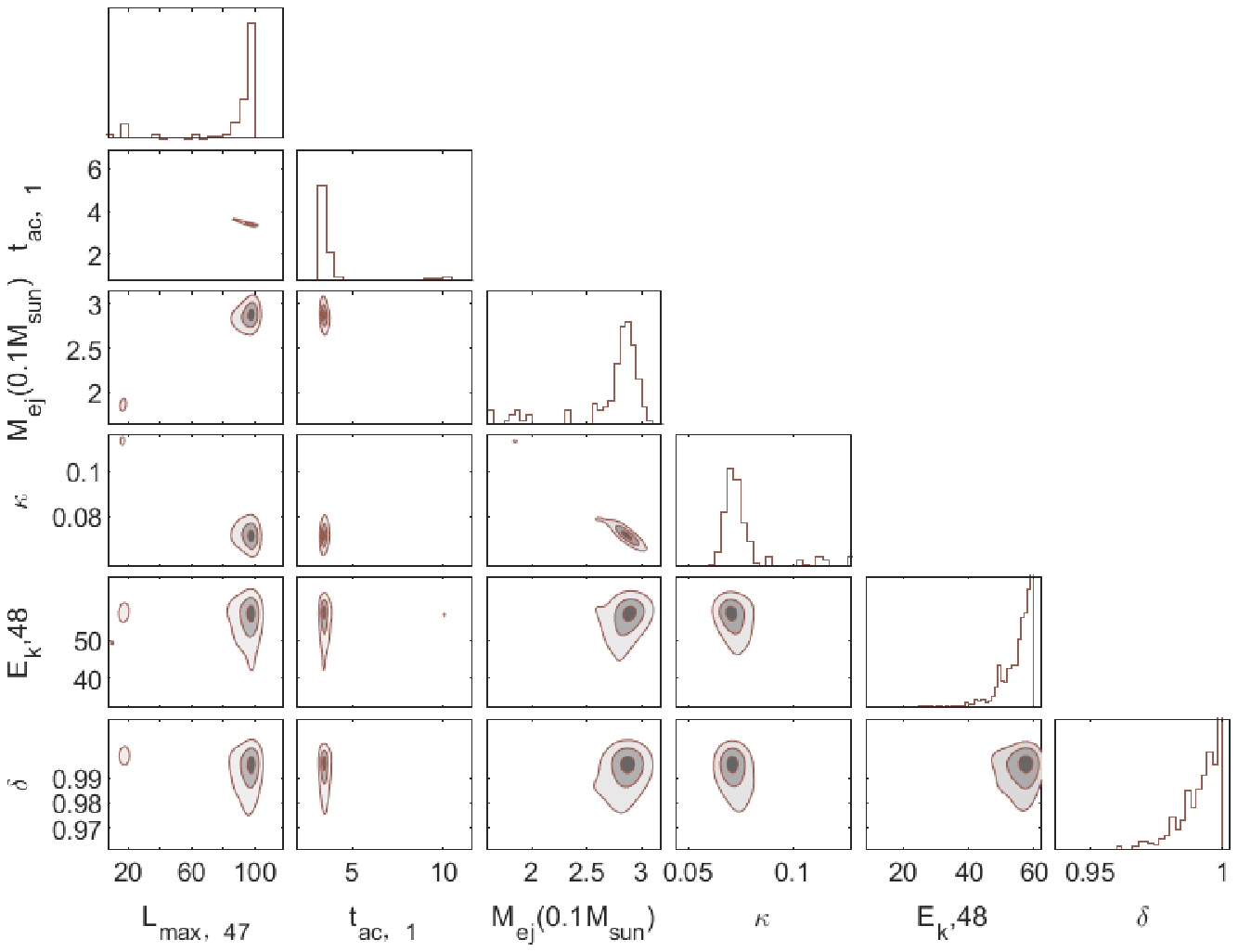}
    \caption{Observational constraints on the model parameters for the spinning-down NS (top) and fallback accretion (bottom) models.}
    \label{fig:C}
\end{figure}

 \begin{table}
 \centering
 \caption{Parameter Values}
 \begin{tabular}{cccc}
 \hline
 \hline
 Parameters & Units & Spinning-Down NS & Fallback Accretion\\
 \hline
 $L_{\rm ce,i}$&$\rm erg~s^{-1}$& $9.43_{-0.87}^{+0.44}\times10^{47}$  & $9.54_{-2.40}^{+0.36}\times10^{48}$ \\
 $t_{\rm sd}~{\rm  or}~ t_{\rm ac}$&$\rm s$          & $4.34_{-0.10}^{+0.23}\times10^{2}$           & $34.40_{-0.78}^{+6.70}$  \\
 $\rm M_{\rm ej}$&$\rm M_{\odot}$            & $0.35_{-0.02}^{+0.02}$               &
 $0.28_{-0.02}^{+0.01}$  \\
 $\kappa$&$\rm  cm^2g^{-1}$          &
 $0.06_{-0.01}^{+0.01}$              &
 $0.07_{-0.01}^{+0.01}$ \\
 $E_{\rm k}$&$\rm 10^{49}erg$        &
 $4.32_{-1.88}^{+1.20}$        &
 $5.62_{-0.67}^{+0.29}$ \\
 $\delta$       &      $-$               & $0.96_{-0.06}^{+0.03}$                        & $0.99_{-0.01}^{+0.00}$ \\
 \hline
 \end{tabular}
 \label{parameters}
\end{table}

SN 2019bkc was found to be associated with a galaxy group at redshift $\sim 0.02$ corresponding to a distance of 89.1 Mpc. This determines its peak luminosity to be around $10^{42}\rm erg ~s^{-1}$, which is comparable to that of AT 2017gfo and difficult to be accounted for by a radioactive power. On the one hand, as discussed for AT 2017gfo in \cite{LiSZ2018ApJ...861L..12L}, if the SN 2019bkc emission is purely powered by the decays of r-process elements, then its ejecta mass should be as high as $\sim\rm 1.0M_{\odot}$ by according to Equation (\ref{Lrpro}). Simultaneously, the opacity is required to be as low as $\sim 0.03\rm cm^2g^{-1}$, in order to explain the rise time of the emission on the order of a few days by the photon diffusion timescale as $t_{\rm d}=(3\kappa M_{\rm ej }/4\pi v c)^{1/2}$. However, in contrast, the typical ejecta masses due to the NS-NS or NS-BH mergers are only around $M_{\rm ej}=10^{-4}-10^{-2}\rm M_\odot$ \citep{Hotokezaka2013PhysRevD.87.024001} and the corresponding opacity is expected to reach $\sim 10-100\rm cm^2g^{-1}$ due to the synthesis of lanthanides \citep{tanaka2013ApJ...775..113T}. Therefore, the SN 2019bkc emission cannot be powered by the decays of r-process elements. On the other hand, SN 2019bkc is also unlikely to be purely powered by the decay of $^{56}$Ni. This is not only because of the unreasonably high requirement on the mass of $^{56}$Ni, which is comparable to or even higher than the total mass of the ejecta, but also because the decline time of the emission is actually shorter than the lifetime of  $^{56}$Ni. Therefore, it is reasonable and necessary to invoke a central engine for explaining the emission characteristics of SN 2019bkc, which has also been previously suggested by \cite{Yoshida2019RNAAS...3..112Y}.

\begin{figure}
      \centering
      \includegraphics[scale=0.45]{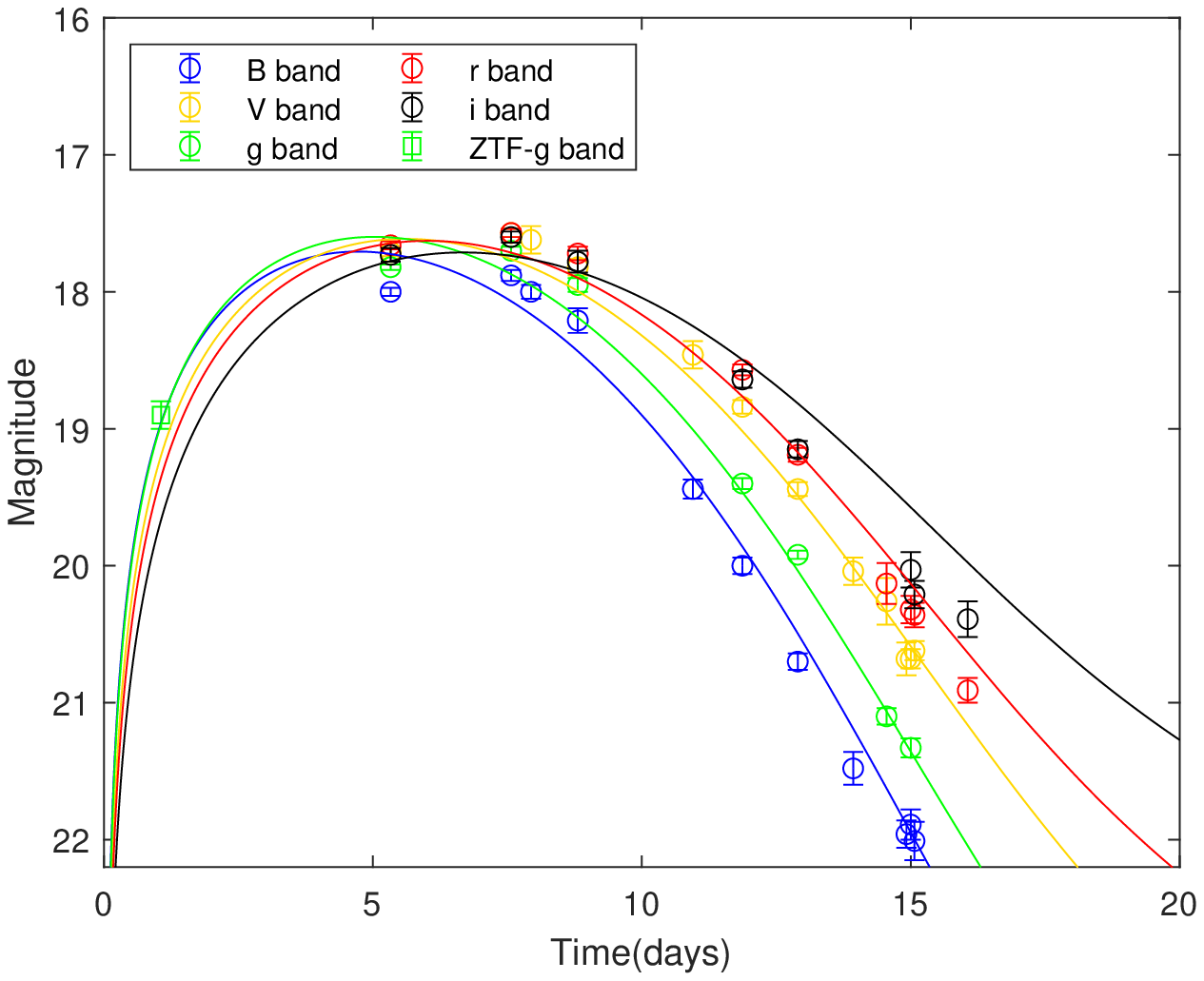}
      \includegraphics[scale=0.45]{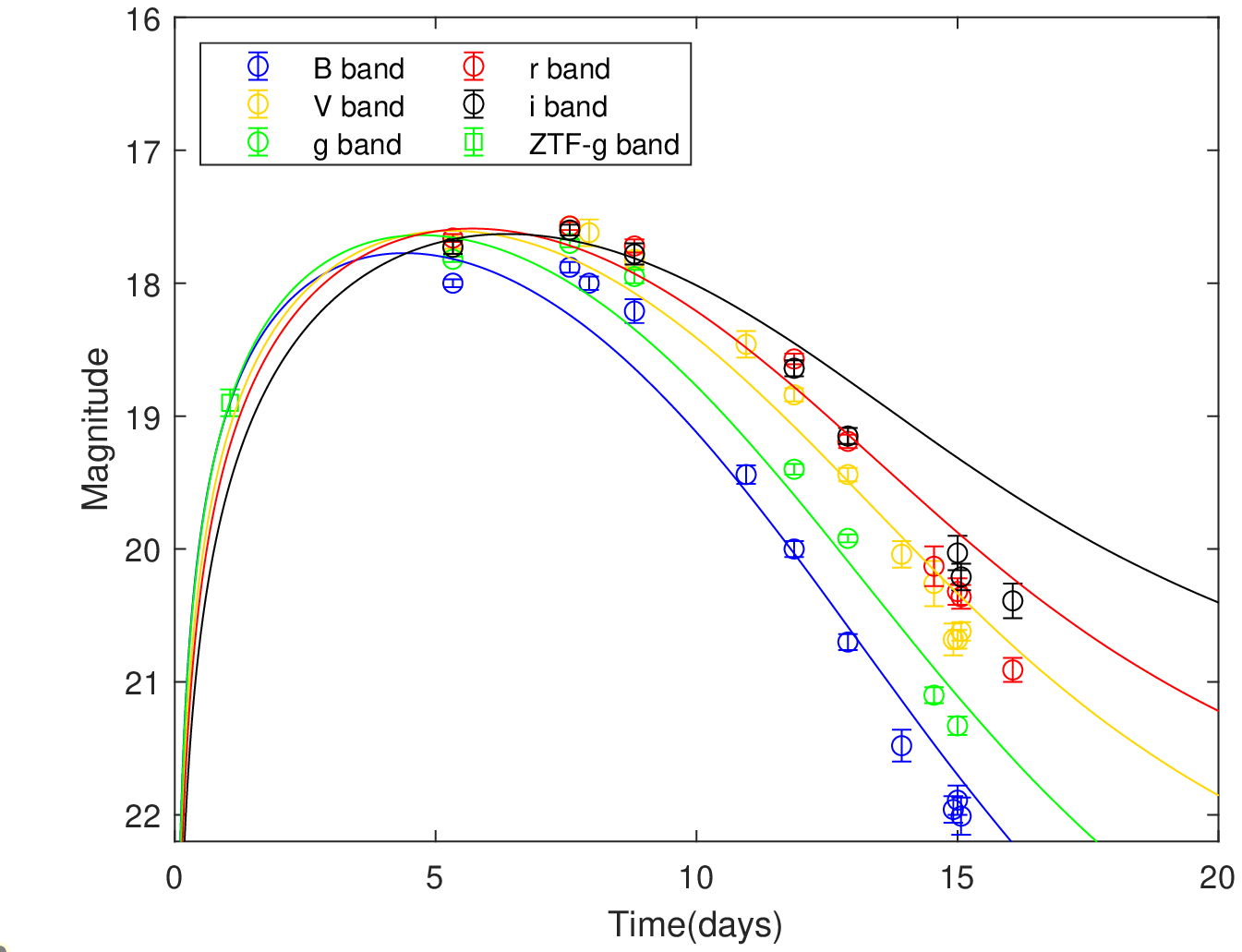}
      \caption{Fittings to the multi-color light curves of SN 2019bkc with the spinning-down NS model (left) and the fallback accretion model (right) for the best-fit parameters. The data are taken from \citep{Chen_2020} and \citep{Prentice2020A&A...635A.186P}. The time origin is set at 2019-02-26 04:48:00 UT (JD 2458540.70).}
      \label{fig:mag}
  \end{figure}

Then, by confronting the engine model with the multi-color observational data of SN 2019bkc, we constrain the model parameters by using the Markov Chain Monte Carlo (MCMC) method \citep{GW2010CAMCS...5...65G}. After 5000 steps with 40 ``walkers", we present the parameter constraints in Fig. \ref{fig:C} for both models of a spinning-down NS and a fallback accretion. The constrained parameter values are listed in Table. \ref{parameters}. Here, besides the central engine, a radioactive power of an amount of 0.002$\rm M_{\odot}$ $^{56}$Ni is still considered, which is necessary for explaining the three late points in $i$ band \citep{Chen_2020}. For the best-fit parameters, we present the fittings of multi-color light curves of SN 2019bkc in Fig. \ref{fig:mag}. It is showed that both models can in principle be consistent with the observational data. The primary deviation of the models from the data appears in the $i$ band. The later the time is, the more serious the deviation will be. This is obviously due to the fact that the de facto emission spectrum can deviate from the black-body spectrum more and more, as the ejecta gradually becomes transparent. This complexity of the emission spectra can be clearly seen from the almost overlap between the $r$ and $i$ data.

Based on the fitting results, if the engine is a spinning-down NS, then we can derive the initial spin period and the dipole magnetic field of the NS to be
\begin{eqnarray}
P_{\rm i}=14.9 \xi^{1/2}L_{\rm sd,i,47}^{-1/2}t_{\rm sd,3}^{-1/2}\rm ms=7.4\xi^{1/2}\rm ms,
\end{eqnarray}
and
\begin{eqnarray}
B_{\rm p}=2.2\times10^{16}\xi^{1/2}L_{\rm sd,i,47}^{-1/2}t_{\rm sd,3}^{-1}\rm G=1.6\times10^{16}\xi^{1/2}\rm G,
\end{eqnarray}
according to the expressions of $\xi L_{\rm sd,i}=9.4\times10^{47}\rm erg~s^{-1}$ and $t_{\rm sd}=4.3\times10^{2}\rm s$.
This result indicates the post-explosion NS is a millisecond magnetar as expected if $\xi \sim \mathcal O(0.1)$, which is well consistent with the previous results found in \cite{Yu_2015} for the PS1 transients. Here the value of $\xi$ could be much smaller than 1 because of the following reasons. (i) Drawing lessons from gamma-ray busts and superluminous supernovae, the wind driven by a millisecond magnetar could be highly anisotropic. The majority energy could be collimated within a small cone around the rotational axis and would not influence the thermal emission of the isotropic ejecta. (ii) The energy released from the isotropic wind component could be partially reflected back into the magnetar wind \citep{Metzger2014MNRAS.439.3916M}. (iii) The energy finally injected into the ejecta can only be absorbed and thermalized in the ejecta with a limited efficiency, which is specifically determined by the emission spectrum of the magnetar wind and the energy-dependent opacity of the ejecta \citep{Yu2019ApJ...877L..21Y}.

Alternatively, if the engine is a fallback accretion, the maximum accretion rate and the total mass of the fallback material can be estimated to
\begin{eqnarray}
\dot{M}_{\rm ac,\max}=L_{\rm ac,\max}/c^2=5.3\times10^{-6}\xi^{-1}\rm M_{\odot}s^{-1},
\end{eqnarray}
and
\begin{eqnarray}
M_{\rm f}\sim\dot{M}_{\rm ac,\max}t_{\rm ac} =1.8\times10^{-5}\xi^{-1}\rm M_{\odot}.
\end{eqnarray}
Meanwhile, the density of the fallback material before it falls back can be estimated to
\begin{eqnarray}
\rho_{\rm f}\sim {(G t_{\rm ac}^2)^{-1}}=1.3\times10^{4}\rm g~cm^{-3},
\end{eqnarray}
since the accretion timescale can be roughly given by the freefall timescale as $t_{\rm ac}\sim (G\rho_{\rm f})^{-1/2}$. By combining the values of $M_{\rm f}$ and $\rho_{\rm f}$, we can further obtain the length-scale of the fallback material as
\begin{eqnarray}
R_{\rm f}\sim \left(\frac{3M_{\rm f}}{4\pi \rho_{\rm f}}\right)^{1/3}\sim 1.9\times10^8 \xi^{-1/3}\rm cm,
\end{eqnarray}
which is just smaller than and somewhat comparable to the typical radius $\sim 10^4$ km of white dwarfs \citep{1983bhwd.book.....S}.

\section{Conclusion and Discussions}
\label{sect:discussion}

In this paper, we demonstrate that the fast-evolving luminous emission of SN 2019bkc can be well explained by an explosion that ejects a mass of $\sim\rm0.3M_{\odot}$ and is lastingly powered by a central engine.

The progenitor of SN 2019bkc is unlikely to be a massive star, because of the deficiency of hydrogen and oxygen features in the spectra and its un-association with a host galaxy. Therefore, the ultra-stripped SN model \citep{Tauris2013ApJ...778L..23T} and the SN shock breakout model are somewhat disfavored. Then, as an alternative natural consideration, the hostlessness of SN 2019bkc could be explained by a NS-NS binary progenitor, the merger of which could lead to the formation of a massive millisecond magnetar. However, this scenario is seriously challenged and even can be ruled out by the required very high mass of the ejecta, even though the ejecta mass can be somewhat increased by a long-lived post-merger NS \citep{Radice2018ApJ...869L..35R}.

In comparison, a WD-involved progenitor could be a better choice. Nevertheless, the requirements on the central engine and the ejecta mass still rule out the possibility that SN 2019bkc is a SN ``.Ia" explosion due to helium-shell detonations on the surface of a sub-Chandrasekhar mass WD \citep{Bildsten2007ApJ...662L..95B,Shen2010ApJ...715..767S}. Alternatively, SN 2019bkc could originate from the collapse/disruption of a WD, specifically, (i) an accretion-induced collapse of a WD \citep{Canal1976A&A....46..229C,Ergma1976AcA....26...69E,Dessart_2006,Yu2019ApJ...877L..21Y,Yu2019ApJ...870L..23Y}, (ii) a merger of a WD with a NS or a stellar-mass black hole \cite[BH; ][]{Metzger2012MNRAS.419..827M,McBrien2019ApJ...885L..23M}, and (iii) the tidal disruption of a WD by a BH of an intermediate mass of $\sim 10^{2.5}\rm M_{\odot}$ \citep{Kawana2020ApJ...890L..26K}. Fairly speaking, the last scenario could be most likely to generate a sufficiently heavy ejecta. However, it is not sure whether such a system can exist far away from galaxies. In view of this, the existence of a NS in the binary could still be very helpful for understanding the hostlessness of SN 2019bkc, due to the possible high kick velocity of the NS that can lead the binary system to depart from the center of the host before the final merger happens.
Therefore, the merger of a WD and a NS could be the most promising origin of SN 2019bkc. After the merger, the remnant NS can have been accelerated to spin at a near-Keplerian frequency due to the accretion of material from the companion WD. Simultaneously, the magnetic field has also been amplified. Then, the transient emission can be primarily powered by the spin-down of this newborn millisecond magnetar. This possible origin of SN 2019bkc makes it similar to another FOT, SN2018kzr, which was discovered by \cite{McBrien2019ApJ...885L..23M}.

Two open questions still exist. (i) It is not sure whether the WD-NS merger model can explain the Ca-rich lines in the later spectra of SN 2019bkc \citep{Prentice2020A&A...635A.186P}, in view of that very different abundances of $^{40}$Ca have been obtained in different simulations of WD-NS mergers.
For example, \cite{Margalit2016MNRAS.461.1154M} found that $^{40}$Ca is the most abundant isotopes in a merger of a NS and a helium WD, the mass of which can be not much lower than 0.01$\rm M_{\odot}$. On the contrary, \cite{Zenati2019MNRAS.486.1805Z} claimed that the ejecta of a WD-NS merger should be calcium-deficient. (ii) The energy injected into the explosion ejecta from the central engine is usually considered to be in the form of high-energy photons such as X-rays that are emitted from the engine outflow. Then, after the ejecta became transparent, this high-energy emission of a luminosity of $10^{40}-10^{41}\rm erg ~s^{-1}$ for months should in principle be detected on the SN 2019bkc's distance. In other words, as argued by \cite{Prentice2020A&A...635A.186P}, the detection of possible X-ray emission can provide a test of the existence of the central engine. However, in fact, a precise prediction of this high-energy emission is actually very dependent on the shock interaction between the engine outflow and the explosion ejecta and also the uncertain shock microphysics. Sometimes, the emission of the engine outflow could be primarily in the UV band and the X-ray emission can be suppressed by an order of magnitude relative to the total power \citep[e.g., see Section 3 in][]{Yu2019ApJ...877L..21Y}.

\begin{acknowledgements}
The authors appreciate the anonymous reviewer for helpful comments, Ping Chen and Subo Dong for providing the data, Ziqian Liu for support in computation, and Jinping Zhu, Aming Chen, Liang-Duan Liu, and Jianfeng Liu for their discussions. This work is supported by the Ministry of Science and Technology of the People's Republic of China (2020SKA0120300) and the
National Natural Science Foundation of China (Grant Nos. 11822302 and 11833003).

\end{acknowledgements}

\bibliography{bibtex.bbl}

\label{lastpage}

\end{document}